\documentclass[12pt]{article}
\usepackage{mathptmx}
\usepackage{cite} 
\usepackage[scaled=0.9]{helvet}
\usepackage{graphicx}
\usepackage{amsmath}

\usepackage{url}  
\usepackage{array}

\evensidemargin\oddsidemargin
\parskip=\medskipamount

\makeatletter
\long\def\@makecaption#1#2{%
  \vskip\abovecaptionskip
  \sbox\@tempboxa{\small{\bfseries #1} \  #2}%
  \ifdim \wd\@tempboxa >\hsize
    \small{\bfseries #1} \  #2\par
  \else
    \global \@minipagefalse
    \hb@xt@\hsize{\hfil\box\@tempboxa\hfil}%
  \fi
  \vskip\belowcaptionskip}
\renewcommand\section{\@startsection {section}{1}{\z@}%
      {-3.25ex\@plus -1ex \@minus -.2ex}%
      {1ex \@plus .2ex}%
      {\normalfont\large\sffamily\bfseries}}
\renewcommand\subsection{\@startsection{subsection}{2}{\z@}%
      {-3ex\@plus -1ex \@minus -.2ex}%
      {0.5ex \@plus .2ex}%
      {\normalfont\normalsize\sffamily\bfseries}}
\renewcommand\subsubsection{\@startsection{subsubsection}{3}{\z@}%
      {-3ex\@plus -1ex \@minus -.2ex}%
      {0.25ex \@plus .2ex}%
      {\normalfont\normalsize\sffamily\bfseries}}
\renewcommand\paragraph{\@startsection{paragraph}{4}{\z@}%
      {3ex \@plus1ex \@minus.2ex}%
      {-1em}%
      {\normalfont\normalsize\sffamily\bfseries}}
\renewcommand\subparagraph{\@startsection{subparagraph}{5}{\z@}%
      {1ex \@plus.5ex \@minus .2ex}%
      {-1em}%
      {\normalfont\normalsize\sffamily\bfseries}}
\makeatother

\def\qsqmax{q^2_\mathrm{max}}
\def\sth{s_\mathrm{th}}
\def\mbstar{m_{B^*}}
\def\modvub{|V_{ub}|}
\def\n#1e#2n{#1\times10^{#2}}
\def\mi{\mathrm{i}}

\def\gev{\,\mathrm{GeV}}
\def\mev{\,\mathrm{MeV}}
\def\d{\mathrm{d}} 
\def\curlyf{\mathcal{F}}

\def\vubresult{(3.90\pm0.32) \times 10^{-3}}

\begin{document}
\begin{flushright}
SHEP--0710
\end{flushright}

\begin{center}\Large\bfseries\sffamily
$\modvub$ from Exclusive Semileptonic $B\to\pi$ Decays
\end{center}

\begin{center}
\textbf{\textsf{Jonathan M Flynn${}^\mathrm{a}$ and Juan
  Nieves${}^\mathrm{b}$}}\\[2ex]
${}^\mathrm{a}$School of Physics and Astronomy, University of
  Southampton\\
  Highfield, Southampton SO17~1BJ, UK\\
${}^\mathrm{b}$Departamento de F\'isica At\'omica, Molecular y
  Nuclear, Universidad de Granada,\\
  E--18071 Granada, Spain
\end{center}
\medskip

\begin{quote}
\begin{center}\textbf{\textsf{Abstract}}\end{center}
We use Omn\`es representations of the form factors $f_+$ and $f_0$ for
exclusive semileptonic $B\to\pi$ decays, paying special attention to
the treatment of the $B^*$ pole and its effect on $f_+$. We apply them
to combine experimental partial branching fraction information with
theoretical calculations of both form factors to extract $\modvub$.
The precision we achieve is competitive with the inclusive
determination and we do not find a significant discrepancy between our
result, $\modvub=(3.90\pm0.32\pm0.18)\times10^{-3}$, and the inclusive
world average value,
$(4.45\pm0.20\pm0.26)\times10^{-3}$~\cite{hfag:2006bi}.
\end{quote}

\section{Introduction}

The magnitude of the Cabibbo-Kobayashi-Maskawa matrix element $V_{ub}$
can be determined from both inclusive and exclusive semileptonic $B$
meson decays. There has been a recent dramatic improvement in the
quality of the experimental data for the exclusive
decays~\cite{Athar:2003yg,Hokuue:2006nr,Aubert:2005cd,Aubert:2006ry,%
Aubert:2006px}, coupled with the appearance of the first dynamical
lattice QCD and improved lightcone sumrule calculations of the
relevant form factors~\cite{Dalgic:2006dt,Okamoto:2004xg,%
Okamoto:2005zg,Mackenzie:2005wu,VandeWater:2006aa,LCSR_04_BZ}.
Dispersive approaches were combined with lattice results
in~\cite{Lellouch:1995yv} and with leading order heavy meson chiral
perturbation theory and perturbative QCD inputs
in~\cite{Burdman:1996kr}. The appearance of the first partial
branching fraction measurements for $B\to\pi l\nu$~\cite{Athar:2003yg}
made it possible~\cite{Fukunaga:2004zz} to combine dispersive
constraints with experimental differential decay rate information and
theoretical calculations of both form factors in limited regions of
$q^2$ in order to improve the determination of $\modvub$.
In~\cite{Arnesen:2005ez} it was shown that the quality of
the inputs now makes it possible for the exclusive determination to
compete in precision with the inclusive
one\footnote{See~\cite{Stewart:ckm2006} for updates of the fits
in~\cite{Arnesen:2005ez}.}. Thus the compatibility of the two
determinations becomes an interesting issue.

To perform the exclusive $\modvub$ extraction one needs a
model-independent parametrisation of the form factors.
In~\cite{Arnesen:2005ez} a parametrisation inspired by dispersive
bounds calculations was used. An alternative simple parametrisation
using a multiply-subtracted Omn\`es representation for $f_+$, based on
unitarity and analyticity properties, was employed
in~\cite{Flynn:2006vr}. A shortcoming in the treatment of the $B^*$
was pointed out in~\cite{Ball:2006jz}. In this letter we have
addressed this by improving the treatment of the $B^*$ within the
Omn\`es framework. We have also incorporated the scalar form factor
$f_0$ in a simultaneous analysis and examined the possible effects of
correlations among lattice inputs. Finally, we have taken advantage of
new experimental data from the BaBar $12$-bin untagged
analysis~\cite{Aubert:2006px}.

The outcome is that the precision achieved for $\modvub$ is indeed
competitive with the inclusive determination and that we do not find a
significant discrepancy between our result,
$\modvub=(3.90\pm0.32\pm0.18)\times10^{-3}$, and the inclusive world
average value, $(4.45\pm0.20\pm0.26)\times10^{-3}$~\cite{hfag:2006bi}.

\section{Omn\`es Parametrisations}

In our previous work~\cite{Omnes_01,Flynn:2006vr} with the Omn\`es
parametrisation~\cite{omnes,mushkelishvili} for the form factor
$f_+(q^2)$, we treated the $B^*$ as a bound state and took the $B\pi$
elastic scattering phase shift to be $\pi$ at threshold,
$\sth=(m_B+m_\pi)^2$. By using multiple subtractions and approximating
the phase shift by $\pi$ from $\sth$ to infinity, this led to a
parametrisation:
\begin{equation}
\label{eq:omnes-sth}
f_+(q^2) = \frac1{\sth-q^2}
 \prod_{i=0}^n \left[ f_+(s_i) \big(\sth-s_i)
 \right]^{\alpha_i(q^2)},
\end{equation}
with $n+1$ subtractions at $q^2 \in \{s_0, s_1,\ldots,s_n\}$, below
threshold (the $\alpha_i(q^2)$ are defined in
equation~(\ref{eq:alpha}) below). This parametrisation requires as
input only the form factor values $\{f_+(q_i^2)\}$ at $n+1$ positions
$q_i^2$.

Using this parametrisation in a combined fit to experimental data and
theoretical form-factor calculations (lattice QCD and lightcone
sumrules) allows an extraction of $\modvub$ with precision competitive
to the inclusive determination. This parametrisation and others were
compared in reference~\cite{Ball:2006jz} where the form factor $f_+$
was determined by fitting BaBar experimental partial branching
fraction data in $12$ bins~\cite{Aubert:2006fv,Aubert:2006ry} and
using $\modvub$ determined from Unitarity Triangle fits. Good
agreement was found between the Omn\`es parametrisation of
equation~(\ref{eq:omnes-sth}) and parametrisations using
\begin{equation}
\label{eq:ze}
f_+(q^2) =
 \frac1{P(q^2)\phi(q^2,t_0)}\sum_{n=0}^\infty a_n\, z(q^2,t_0)^n
\end{equation}
for two choices of $t_0$. The coefficients $a_n$ satisfy the
dispersive constraint $\sum_n a_n^2 \leq 1$~\cite{Arnesen:2005ez}.
Expressions for $P$ and $\phi$ can be found in~\cite{Arnesen:2005ez}.
When we use the parametrisation in equation~(\ref{eq:ze}), we will set
$t_0 = \sth(1-\sqrt{1-\qsqmax/\sth})$, which is the `preferred
choice', labelled BGLa, in~\cite{Ball:2006jz} (this choice for $t_0$
ensures that $|z|\leq0.3$ for $0\leq q^2\leq\qsqmax$). We will refer
to the parametrisation using this functional form as the $z$-expansion
or ZE below.

\begin{figure}
\begin{center}
\includegraphics[width=0.48\hsize]{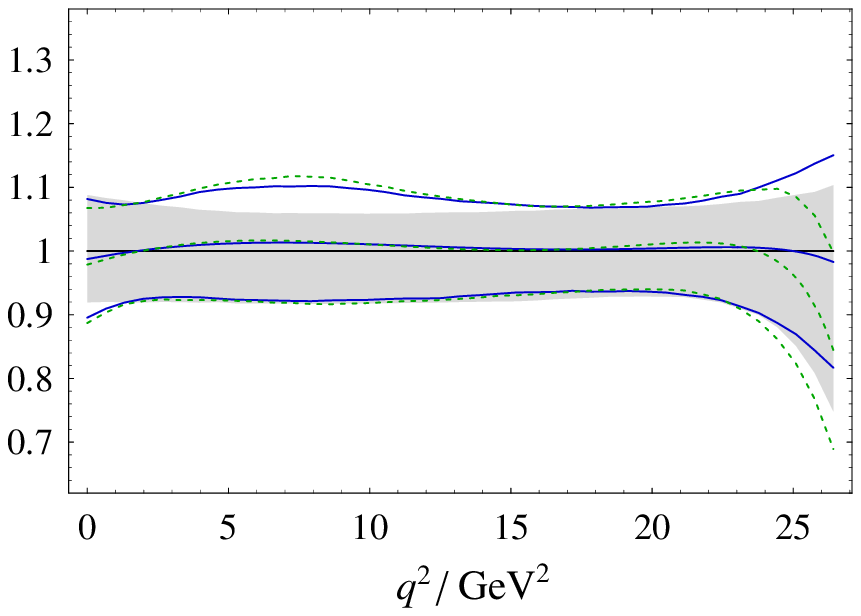}
\hfill
\includegraphics[width=0.48\hsize]{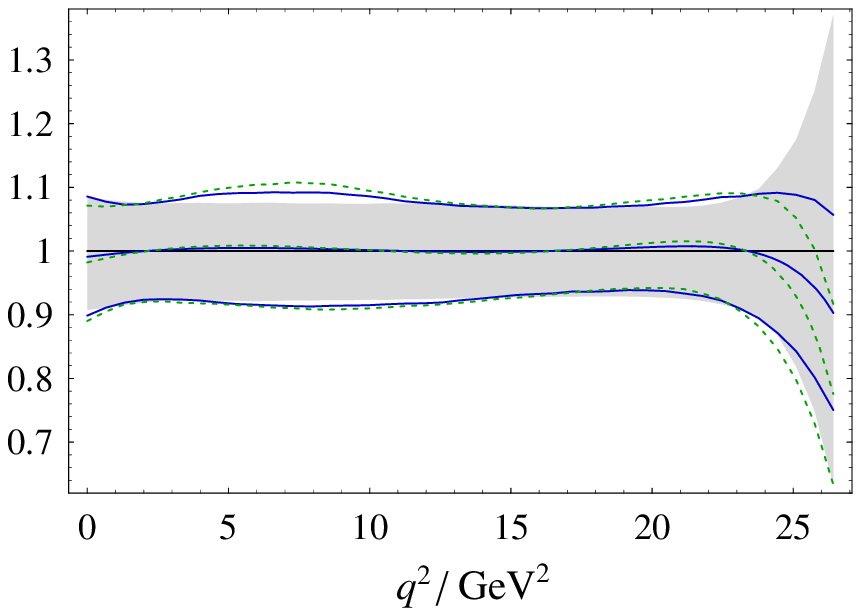}
\end{center}
\caption{Comparison of fits to $f_+(q^2)$ using Omn\`es or ZE
  parametrisations. Each fit is plotted with its own error bands, but
  normalised by the central fit for the ZE. Thus, the horizontal line
  at $1$ and the grey band show the ZE fit with its $68\%$ statistical
  error band, the solid blue lines indicate the Omn\`es fit of
  equation~(\ref{eq:omnes-mbstar}) and the green dashed lines show the
  Omn\`es fit of equation~(\ref{eq:omnes-sth}). The left-hand plot
  uses a ZE fit with three parameters and thus can be compared to
  figure~$2$ in~\cite{Ball:2006jz}, while the right-hand plot uses a
  four-parameter ZE fit. All Omn\`es fits use four parameters (four
  subtraction points).}
\label{fig:compare}
\end{figure}

Fits for $f_+(q^2)$ using the Omn\`es and ZE parametrisations deviated
from each other by a few per cent only in the largest $q^2$ region,
close to $\qsqmax$, which has little influence on the decay width and
$\modvub$ (see figure~$2$ in~\cite{Ball:2006jz}). This is also the
region where there is no theoretical information on the form factors.
In figure~\ref{fig:compare} we show a similar comparison, including
bands showing statistical fluctuations arising from the fits. We have
fitted the same dataset as in~\cite{Flynn:2006vr}, but replacing the
$5$-bin BaBar untagged analysis~\cite{Aubert:2005cd} with the updated
$12$-bin results from~\cite{Aubert:2006px}. The ZE fit has been
performed truncating the power series in equation~(\ref{eq:ze}) at
$n=2$, for comparison with figure~$2$ in~\cite{Ball:2006jz}, or $n=3$,
so that all fits have the same number of parameters. The green dashed
lines show the Omn\`es fit using equation~(\ref{eq:omnes-sth}). The
plot shows that once fluctuations are taken into account the
differences are not significant.

Nevertheless, we show here that by treating the $B^*$ explicitly as a
pole of the form factor, we can understand and reduce the small
deviation in the central fits at large $q^2$. This is illustrated by
the solid blue lines in figure~\ref{fig:compare}. We achieve this
without altering the main results obtained for $\modvub$ and $f_+$ in
the $q^2$ region where theoretically calculated values lie. The new
parametrisation, shown below in equation~(\ref{eq:omnes-mbstar}), is
obtained from equation~(\ref{eq:omnes-sth}) by replacing $\sth$ with
$\mbstar^2$. As before, the parametrisation relies only on very
general properties of analyticity and unitarity and so, although
simple, is well-founded.

To obtain the new parametrisation we observe that, if $f_+(q^2)$ has a
pole at $q^2=m_{B^*}^2$, then $\curlyf(q^2) \equiv (\mbstar^2 -
q^2)f_+(q^2)$ has no poles and satisfies
\begin{equation}
\frac{\curlyf(s+i\epsilon)}{\curlyf(s-i\epsilon)} =
 \exp\big({2\mi \delta_{1/2,1}(s)}\big),\qquad\mbox{$s\geq\sth$}
\end{equation}
where $\delta_{IJ}$ is the phase-shift for elastic $\pi B \to \pi B$
scattering in the isospin $I$ and total angular momentum $J$ channel.
This is because $f_+$ satisfies a similar equation as required by
Watson's theorem~\cite{wa54} and we have multiplied it by a real
function. An $(n{+}1)$-subtracted Omn\`es representation can now be
written for $\curlyf(q^2)$, with $q^2<\sth$, which reads:
\begin{align}
\curlyf(q^2) &=
  \bigg(\prod_{i=0}^n\left[\curlyf(s_i)\right]^{\alpha_i(q^2)}\bigg)
  \exp\bigg\{I_\delta(q^2;\,s_0,\ldots,s_n)
             \prod_{j=0}^n(q^2-s_j)
      \bigg\}, \label{eq:omnes} \\ 
I_\delta(q^2;\, s_0,\ldots,s_n) &=
  \frac1{\pi}\int_{\sth}^{+\infty}
  \frac{\d s}{(s-s_0)\cdots(s-s_n)}\,\frac{\delta_{1/2,1}(s)}{s-q^2},
\label{eq:phase-integral}\\
\alpha_i(s) &\equiv \prod_{j=0, j\neq i}^n
        \frac{s-s_j}{s_i-s_j},\qquad
\alpha_i(s_j)=\delta_{ij},\qquad
\sum_{i=0}^n \alpha_i(s) = 1.
\label{eq:alpha}
\end{align}
This representation requires as input the phase shift
$\delta_{1/2,1}(s)$ plus the values $\{\curlyf(s_i)\}$ at $n+1$
positions $\{s_i\}$ below the $\pi B$ threshold. For sufficiently many
subtractions, we can approximate $\delta_{1/2,1}(s)$ by zero above
threshold (see appendix~\ref{appx:omnes}). In this case we obtain,
\begin{equation}
\label{eq:omnes-mbstar}
f_+(q^2) = \frac1{\mbstar^2-q^2}
 \prod_{i=0}^n \left[ f_+(s_i) \big(\mbstar^2-s_i)
 \right]^{\alpha_i(q^2)}.
\end{equation}
This amounts to finding an interpolating polynomial for
$\ln\curlyf(q^2)=\ln[(\mbstar^2-q^2)f_+(q^2)]$ passing through the
points $\curlyf(s_i)=(\mbstar^2-s_i)f_+(s_i)$. Similarly, our earlier
parametrisation in equation~(\ref{eq:omnes-sth}) used an interpolating
polynomial for $\ln[(\sth-q^2)f_+(q^2)]$. While one could always
propose a parametrisation using an interpolating polynomial for
$\ln[g(q^2)f_+(q^2)]$ for a suitable function $g(q^2)$, the derivation
using the Omn\`es representation shows that taking
$g(q^2)=\mbstar^2-q^2$ (and equally $\sth-q^2$) is physically
motivated. From here onwards we will use
equation~(\ref{eq:omnes-mbstar}) as our preferred parametrisation for
$f_+$.

When using our parametrisation in the extraction of $\modvub$, we make
$4$ subtractions. This is sufficient to justify using no information
about the phase shift beyond its value at $\sth$. To check this, we
have put in a model for the $B\pi$ phase shift~\cite{Omnes_01} and
confirmed that induced changes in our results are much smaller than
the fluctuations produced by the errors in our inputs. This can be
understood because with four evenly-spaced subtractions at
$\{0,1/3,2/3,1\}\qsqmax$, the factor $\exp\big[I_\delta\times
\prod_{j=0}^n(q^2-s_j)\big]$ in equation~(\ref{eq:omnes}) given by
this model deviates from unity by no more than $\n6e-4n$ for $0\leq
q^2\leq\qsqmax$ (and, of course, is unity at each subtraction point).

Since $f_+$ and and the scalar form factor $f_0$ satisfy the
constraint $f_+(0)=f_0(0)$ we will combine theoretical inputs for
$f_+$ and $f_0$ with experimental $B\to\pi l\nu$ partial branching
fraction information to check the effect on the extracted value of
$\modvub$. We will investigate the effect of using the $f_+$
information alone or using both form factors.

For the scalar form factor $f_0$ there is no pole below threshold, so
that we will use an Omn\`es formula like equation~(\ref{eq:omnes}) for
$f_0(q^2)$, with $\curlyf\to f_0$ and $\delta_{1/2,1} \to
\delta_{1/2,0}$. For sufficiently many subtractions, we can
approximate $\delta_{1/2,0}$ by zero above threshold. Our recent
analysis of the scalar form factor~\cite{Flynn:2007ki} for $B\to\pi$
decays suggested the existence of a resonance with mass around
$5.6\gev$. This could be incorporated in an Omn\`es parametrisation
like that of equation~(\ref{eq:omnes-mbstar}), but (as we have
confirmed) has negligible effect on $\modvub$ and $f_+$ in our fit,
producing only a small increase of around $7\%$ in the value of $f_0$
close to $\qsqmax$.

\section{Application to $\modvub$}

We have used experimental data for the partial branching fractions of
$B\to\pi l\nu$ decays in $q^2$ bins from both tagged and untagged
analyses. The tagged analyses from CLEO~\cite{Athar:2003yg},
Belle~\cite{Hokuue:2006nr} and BaBar~\cite{Aubert:2006ry} use three
bins, while BaBar's untagged analysis~\cite{Aubert:2006px} uses
twelve. CLEO and BaBar combine results for neutral and charged
$B$-meson decays using isospin symmetry, while Belle quote separate
values for $B^0\to \pi^- l^+ \nu_l$ and $B^+\to\pi^0 l^+\nu_l$. For
our analysis, for the three-bin data, we have combined the Belle
charged and neutral $B$-meson results and subsequently combined these
with the CLEO and BaBar results. The resulting input values can be
found in table~II of~\cite{Flynn:2006vr}. Since the systematic errors
of the three-bin data are small compared to the statistical ones, we
have ignored correlations in the systematic errors and combined errors
in quadrature. For the $12$-bin BaBar data~\cite{Aubert:2006px},
complete correlation matrices are available in the EPAPS
database~\cite{BaBar:EPAPS} for both statistical and systematic errors
and we have used these in our fits (we used the results corrected for
final state radiation effects). We have assumed no correlation between
the untagged and the tagged analyses.

When computing partial branching fractions, we have used $\tau_{B^0}=
1/\Gamma_\mathrm{Tot} = \n(1.527\pm
0.008)e-12n\,\mathrm{s}$~\cite{hfag:2006bi} for the $B^0$ lifetime.

Since the effects of finite electron and muon masses are beyond
current measurement precision, the experimental results provide
information on the $q^2$ shape of $f_+$. Theoretical calculations
provide information on $f_+$ and $f_0$.

We use the lightcone sumrule (LCSR) result $f_+(0)=f_0(0) =
0.258\pm0.031$~\cite{LCSR_04_BZ} and lattice QCD results from
dynamical simulations at larger $q^2$ from HPQCD~\cite{Dalgic:2006dt}
and
FNAL-MILC~\cite{Okamoto:2004xg,Okamoto:2005zg,Mackenzie:2005wu,VandeWater:2006aa}.
The FNAL-MILC
results~\cite{Okamoto:2004xg,Okamoto:2005zg,Mackenzie:2005wu,VandeWater:2006aa}
are still preliminary. Therefore we use the three $f_+(q^2)$ values quoted
in~\cite{Arnesen:2005ez} and read off three values for $f_0(q^2)$ at
the same $q^2$ points from~\cite{Okamoto:2005zg}. These are
\begin{equation}
\begin{aligned}
f_0(15.87\gev^2) &= 0.425\pm0.033 \\
f_0(18.58\gev^2) &= 0.506\pm0.037 \\
f_0(24.09\gev^2) &= 0.800\pm0.067
\end{aligned}
\end{equation}
The errors shown are statistical. A further $11\%$ systematic error
should be added.

We implement the fitting procedure described in~\cite{Flynn:2006vr}
using four evenly-spaced Omn\`es subtraction points at
$\{0,1/3,2/3,1\}\qsqmax$ (with $\chi$-squared function given in
equation~(10) of~\cite{Flynn:2006vr}), with the obvious changes to
incorporate $f_0$. As before, we have assumed that the lattice input
form factor data have independent statistical uncertainties and
fully-correlated systematic errors. We have not assumed correlations
between results for $f_+$ and $f_0$, though we will comment further on
this below. Furthermore, we ignore possible correlations between the
HPQCD and FNAL-MILC lattice inputs. These correlations are unknown and
we showed in~\cite{Flynn:2006vr} that unless they are very strong they
will have little effect on $\modvub$.

The best-fit parameters are
\begin{equation}
\label{eq:best-fit}
\begin{array}{rcl}
\modvub         &=& \vubresult \\
f_+(0)=f_0(0)   &=& 0.226\pm0.022 \\
f_+(\qsqmax/3)  &=& 0.417\pm0.039 \\
f_+(2\qsqmax/3) &=& 0.941\pm0.064 \\ 
f_+(\qsqmax)    &=& 7.29 \pm1.28 \\
f_0(\qsqmax/3)  &=& 0.342\pm0.053 \\
f_0(2\qsqmax/3) &=& 0.508\pm0.040 \\ 
f_0(\qsqmax)    &=& 1.09\pm0.21
\end{array}
\end{equation}
The fit has $\chi^2/\mathrm{dof} = 0.62$ for $28$ degrees of freedom,
while the Gaussian correlation matrix can be found in
appendix~\ref{appx:corr}.

In figure~\ref{fig:results} we show the fitted form factors, the
differential decay rate calculated from our fit and the quantities
$\log[(\mbstar^2-q^2)f_+(q^2)/\mbstar^2]$ and $P\phi f_+$ where the
details of the fit and inputs can better be seen. The dashed magenta
curve in the $P\phi f_+$ plot is a cubic polynomial fit in $z$ to the
output from our analysis. We note that the sum of squares of the
coefficients in this polynomial safely satisfies the dispersive
constraint $\sum_n a_n^2 \leq 1$~\cite{Arnesen:2005ez}.
\begin{figure}
\begin{center}
\includegraphics[width=\hsize]{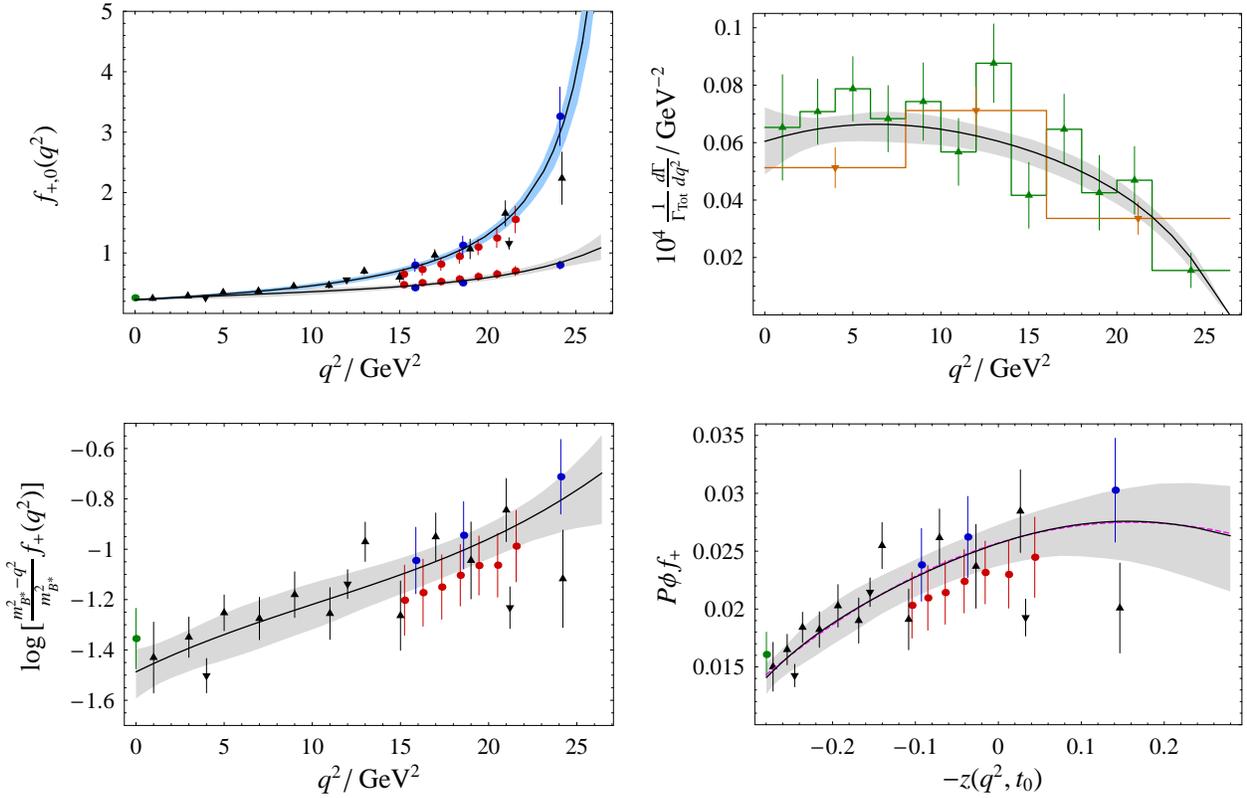}
\end{center}
\caption{Results obtained from the fit to experimental partial
  branching fraction data and theoretical form factor calculations.
  The top left plot shows the two form factors with their error bands,
  the lattice and LCSR input points (dots: green LCSR, red HPQCD, blue
  FNAL-MILC) and `experimental' points (black triangles,
  upward-pointing for tagged and downward pointing for untagged data)
  constructed by plotting at the centre of each bin the constant form
  factor that would reproduce the partial branching fraction in that
  bin. The top right plot shows the differential decay rate together
  with the experimental inputs. The bottom plots provide more details
  of the inputs and fits by showing on the left
  $\log[(\mbstar^2-q^2)f_+(q^2)/\mbstar^2]$ as a function of $q^2$,
  and on the right $P\phi f_+$ as a function of $-z$. The dashed
  magenta curve in the bottom right plot is a cubic polynomial fit in
  $z$ to the Omn\`es curve.}
\label{fig:results}
\end{figure}

Compared to our previous results~\cite{Flynn:2006vr} we find that the
central value of $\modvub$ decreases by $3\%$ compared with an error
of around $8\%$. Similarly, the central values of $f_+(0)$ and
$f_+(\qsqmax)$ move up by around half their errors, while
$f_+(\qsqmax/3)$ increases by an amount comparable with its error. At
$2\qsqmax/3$, in the neighbourhood of which most of the form factor
data is concentrated, there is hardly any change. The result for
$f_0(\qsqmax)$ agrees with that obtained in our recent analysis of the
scalar form factor alone~\cite{Flynn:2007ki}. We make some remarks
on these results:
\begin{itemize}
\item We have checked that the changes in the results for $f_+(0)$,
  $f_+(\qsqmax/3)$ and $\modvub$ stem from using the updated BaBar
  untagged data.
\item We have checked that the change in $f_+(\qsqmax)$, which has
  little effect on the shape of the form factor in the $q^2$ range
  where experimental and theoretical information exists, arises from
  our use of the new Omn\`es parametrisation of
  equation~(\ref{eq:omnes-mbstar}) and reflects the existence of a
  pole in $f_+$ at $q^2=\mbstar^2$.
\item Since we do not know the correlations between the lattice input
  data we have also performed a fit neglecting all correlations in
  these inputs. We find that $\modvub$ increases by an amount
  $0.18\times10^{-3}$, which we will quote as a systematic error in
  our determination. We observe that knowledge of the correlations
  will be needed for more precise determinations of $\modvub$.
\item The inclusion of $f_0$ in the analysis has no visible effect in
  our results for $f_+$ and $\modvub$. This is not surprising given
  that the number of input data affecting $f_+$ is much bigger than
  that affecting $f_0$ and that the parametrisation allows the data to
  determine each form factor independently apart from the constraint
  at $q^2=0$. The covariance matrix given in appendix~\ref{appx:corr}
  shows this freedom, having negligible correlations between $f_+$ and
  $f_0$ at $q^2\neq0$. Correlations linking $f_+$ and $f_0$ in the
  lattice QCD inputs could modify the central values
  in~(\ref{eq:best-fit}) by an amount comparable to their errors as we
  have confirmed by fully-correlating the systematic errors between
  them. As an example, for $\modvub$ we find a central value of
  $4.15\times10^{-3}$. Since we do not know the actual correlation
  information\footnote{It is reasonable to expect correlations not
  only in the systematic error but also in the statistical ones, since
  $f_+$ and $f_0$ are linear combinations of temporal and spatial
  components of vector current matrix elements.} for the lattice data,
  we do not present these numbers.
\item Because of the freedom allowed by the Omn\`es parametrisation of
  $f_+$ and $f_0$, one may wonder whether or not heavy quark symmetry
  (HQS) relations between the form factors at $\qsqmax$ are satisfied.
  Some earlier parametrisations were explicitly constructed to satisfy
  the HQS scaling relation $f_+(\qsqmax)/f_0(\qsqmax) \sim m_B$, for
  example dipole/pole
  forms~\cite{Burford:1995fc,DelDebbio:1997kr,Becirevic:1999kt}, and
  these have been widely used. From our fit we calculate
  \begin{equation}
  \frac1{m_B} \left.\frac{f_+(\qsqmax)}{f_0(\qsqmax)}\right|_{B\pi}
   = 1.3\pm0.4\gev^{-1}
  \end{equation}
  to be compared to the corresponding quantity in $D\to\pi$ exclusive
  semileptonic decays, $1.4\pm0.1\gev^{-1}$ extracted from the
  unquenched lattice QCD results in~\cite{Aubin:2004ej}. This
  agreement is reassuring but our determination of the ratio in
  $B\to\pi$ decays has a further uncertainty of around $10\%$ arising
  from our incomplete knowledge of the correlations in the lattice
  inputs.
\item Heavy quark effective theory in the soft-pion limit
  predicts~\cite{Burdman:1993es},
  \begin{equation}
  f_0(m_B^2) = f_B/f_\pi + \mathcal{O}(1/m_b^2) \approx 1.4(2)
  \end{equation}
  where we have used $f_B=189(27)\mev$~\cite{Hashimoto:2004hn}. Our
  fit for $f_0(\qsqmax)$ in equation~(\ref{eq:best-fit}) is compatible
  within errors.
\item Applying soft collinear effective theory (SCET) to $B\to\pi\pi$
  decays allows a factorisation result to be derived which leads to a
  model-independent extraction of the form factor (multiplied by
  $\modvub$) at $q^2=0$~\cite{vubfplus-fact-2004}. We quote the result
  from our fit,
  \begin{equation}
  \modvub f^+(0) = (8.8\pm0.8)\times10^{-4},
  \end{equation}
  which compares well with $\modvub f^+(0) = \n(7.6\pm1.9)e-4n$ quoted
  in~\cite{Stewart:ckm2006}. This also agrees with the value $\modvub
  f^+(0) = \n(9.1\pm0.7)e-4n$~\cite{Ball:2006jz} obtained by fixing
  $\modvub$ from global CKM unitarity triangle fits and fitting to the
  BaBar $12$-bin data~\cite{Aubert:2006fv}.  
\item We noted above possible effects of correlations in the lattice
  data. Other sources of systematic variation in the result for
  $\modvub$ arising from uncertainties in the theoretical form factor
  inputs at or near $q^2=0$ were considered in~\cite{Flynn:2006vr} and
  shown to be safely covered by the statistical uncertainty.
\end{itemize}

\section{Conclusion}

We have updated our previous analysis of exclusive $B\to\pi$
semileptonic decays, based on Omn\`es dispersion relations. The
principal change is to improve the treatment of the $B^*$ and its
effect on the form factor $f_+$. We have also incorporated the scalar
form factor $f_0$ in a simultaneous analysis and examined the possible
effects of correlations among lattice inputs. Finally, we have taken
advantage of new experimental data from the BaBar $12$-bin untagged
analysis~\cite{Aubert:2006px}. We extract a value
\begin{equation}
\modvub = (3.90\pm0.32\pm0.18)\times10^{-3}.
\end{equation}
The first error above is statistical arising from the chi-squared fit.
The second is a systematic error to account for current partial
knowledge of correlations in the lattice input data. The precision for
$\modvub$ is comparable with that of the inclusive determination and
we do not find a significant discrepancy between our result and the
inclusive world average value,
$(4.45\pm0.20\pm0.26)\times10^{-3}$~\cite{hfag:2006bi}.

Finally we would like to stress that the Omn\`es parametrisation is
physically motivated and simple and provides a robust framework for a
precise exclusive determination of $\modvub$.

\subsubsection*{Acknowledgements}

JMF acknowledges the hospitality of the Departamento de F\'isica
At\'omica, Molecular y Nuclear, Universidad de Granada, MEC support
for estancias de Profesores e investigadores extranjeros en r\'egimen
de a\~no sab\'atico en Espa\~na SAB2005--0163, and PPARC grant
PP/D000211/1. JN acknowledges support from Junta de Andalucia grant
FQM0225 and MEC grant FIS2005--00810. JMF and JN acknowledge support
from the EU Human Resources and Mobility Activity, FLAVIAnet, contract
number MRTN--CT--2006--035482.

\appendix
\section{Choice of $\delta_{IJ}(\sth)$ in the Omn\`es Representation}
\label{appx:omnes}

In this appendix we provide more details on some aspects of the
Omn\`es representation of the form factors. This builds on the
discussion in the appendix of~\cite{Albertus:2005ud}.

The scattering matrix $T$ depends on $\exp(2i\delta)$ and thus one has
the freedom to add factors of $k\pi$ to the phase shift, for integer
$k$, without modifying the $T$ matrix. However, the Omn\`es
representation of the form factor certainly depends on the specific
value of $k$. Indeed adding $k\pi$ to $\delta$ leads to
\begin{equation}
\label{eq:omnes-kpi}
\exp\bigg\{I_{\delta+k\pi}\times
             \prod_{j=0}^n(q^2-s_j)\bigg\}
 = \exp\bigg\{I_\delta\times
             \prod_{j=0}^n(q^2-s_j)\bigg\}
   \left(\frac{\prod_{j=0}^{n}(\sth-s_j)^{\alpha_j(q^2)}}{\sth-q^2}
   \right)^k
\end{equation}
which induces an unphysical $k$th order pole in the form factor at
$\sth$.

Now consider $\curlyf(q^2)=(\mbstar^2-q^2)f_+(q^2)$, which has no
poles in $0\leq q^2\leq\sth$. Its Omn\`es representation should not
induce a pole at $\sth$ and therefore we should set $k=0$ in
equation~(\ref{eq:omnes-kpi}) above. This is equivalent to setting
$\delta_{1/2,1}(\sth)=0$. With enough subtractions, we can then take
$\delta_{1/2,1}(s)=0$ inside the integral because only the region
close to $\sth$ will be important, leading to the result presented in
equation~(\ref{eq:omnes-mbstar}).

This choice for $\delta_{1/2,1}(\sth)$ does not contradict Levinson's
theorem, which fixes only the difference
\begin{equation}
\delta_{1/2,1}(\infty) - \delta_{1/2,1}(\sth) = \pi (n_z-n_p)
\end{equation}
where $n_z$ ($n_p$) is the number of zeros (poles) of the scattering
matrix $T$ on the physical sheet. The usual convention~\cite{MS70} is
to set $\delta_{1/2,1}(\sth)=\pi n_p$ and $\delta_{1/2,1}(\infty) =
\pi n_z$. However, we use a different convention which follows from
the discussion above on the effect of adding multiples of $\pi$ to the
phase shift. Our choice is $\delta_{1/2,1}(\sth) = 0$ which therefore
also implies that $\delta_{1/2,1}(\infty)=0$.

In our previous work~\cite{Omnes_01,Albertus:2005ud,Flynn:2006vr}, we
assumed that $f_+$ had no poles. With the usual convention for
Levinson's theorem that $\delta_{1/2,1}(\sth) = \pi$, we developed a
pole for $f_+$ at $\sth$ which was not discarded since it mimicked the
$B^*$ pole's effects on the form factor because
$m_{B^*}^2\approx\sth$. We already commented on this in the appendix
of~\cite{Albertus:2005ud}.

\section{Correlation Matrix}
\label{appx:corr}
Here we give the correlation matrix of fitted parameters corresponding
to the best-fit parameters in equation~(\ref{eq:best-fit}).
\begin{equation}
\left(
\begin{array}{cccccccc}
1 & -0.43 & -0.91 & -0.81 & -0.58 & -0.04 &  0.00 &  0.01 \\
  &  1    &  0.20 &  0.50 & -0.04 &  0.10 &  0.00 & -0.02 \\
  &       &  1    &  0.76 &  0.61 &  0.02 &  0.00 &  0.00 \\
  &       &       &  1    &  0.36 &  0.05 &  0.00 & -0.01 \\
  &       &       &       &  1    &  0.00 &  0.00 &  0.00 \\
  &       &       &       &       &  1    &  0.32 &  0.83 \\
  &       &       &       &       &       &  1    &  0.22 \\
  &       &       &       &       &       &       &  1
\end{array}
\right)
\end{equation}

\bibliographystyle{elsevier}
\bibliography{omnes2}

\end{document}